\documentclass[%
 reprint,
 amsmath,amssymb,
 aps,
 prl,
]{revtex4-1}

\usepackage{amsmath}
\usepackage{graphicx}% Include figure files
\usepackage{dcolumn}% Align table columns on decimal point
\usepackage{bm}% bold math
\usepackage{xcolor}

\def\n{n}
\def\sss{\scriptscriptstyle\rm}
\def\ks{_{\sss KS}}
\def\dmrg{_{\sss DMRG}}

\def\s{_{\sss S}}
\def\xc{_{\sss XC}}
\def\xctheta{_{\sss XC,\theta}}

\def\H{_{\sss H}}

\def\br{{\bf r}}

\usepackage[encapsulated]{CJK}

\begin{document}
\begin{CJK*}{UTF8}{gbsn}

\title{Kohn-Sham equations as regularizer:\\building prior knowledge into machine-learned physics}% Force line breaks with \\

\author{Li Li (李力)$^{1}$}
\email[email: ]{leeley@google.com}
\author{Stephan Hoyer$^{1}$}
\author{Ryan Pederson$^{2}$}
\author{Ruoxi Sun (孙若溪)$^{1}$}
\author{Ekin D.~Cubuk$^{1}$}
\author{Patrick Riley$^{1}$}
\author{Kieron Burke$^{2,3}$}
\affiliation{
$^1$ Google Research, Mountain View, CA 94043, USA \\
$^2$ Department of Physics and Astronomy, University of California, Irvine, CA 92697, USA \\
$^3$ Department of Chemistry, University of California, Irvine, CA 92697, USA}

\date{\today}% It is always \today, today,

\begin{abstract}
Including prior knowledge is important for effective machine learning models in physics, and is usually achieved by explicitly adding loss terms or constraints on model architectures.  Prior knowledge embedded in the physics computation itself rarely draws attention.
We show that solving the Kohn-Sham equations when training neural networks for the exchange-correlation functional provides an implicit regularization that greatly improves generalization.
Two separations suffice for learning the entire one-dimensional H$_2$ dissociation curve within chemical accuracy, including the strongly correlated region.
Our models also generalize to unseen types of molecules and overcome self-interaction error.
\end{abstract}
\maketitle
\end{CJK*}

Differentiable programming~\cite{baydin2017automatic} is a general paradigm of deep learning, where parameters in the computation flow are trained by gradient-based optimization. Based on the enormous development in automatic differentiation libraries~\cite{jax2018github,flux,paszke2019pytorch,tensorflow2015-whitepaper}, hardware accelerators~\cite{jouppi2017datacenter} and deep learning~\cite{lecun2015deep}, this emerging paradigm is relevant for scientific computing.
It supports extremely strong physics prior knowledge and well-established numerical methods~\cite{Innes2019-qs} and parameterizes the approximation by a neural network, which can approximate any continuous function~\cite{hornik1991approximation}. 
Recent highlights include discretizing partial differential equations~\cite{Bar-Sinai2019-no}, structural optimization~\cite{hoyer2019neural}, sampling equilibrium configurations~\cite{Noe2019-boltzmann}, differentiable molecular dynamics~\cite{jaxmd2019}, differentiable programming tensor networks~\cite{liao2019differentiable}, optimizing basis sets in Hartree-Fock~\cite{Tamayo-Mendoza2018-auto-diff} and variational quantum Monte Carlo~\cite{hermann2020deep,Pfau2019-bm,yang2020deep}.

Density functional theory (DFT), an approach to electronic structure problems, took 
an enormous step forward with the creation of the Kohn-Sham (KS) equations~\cite{kohn1965self}, which greatly improves accuracy from the original DFT~\cite{hohenberg1964inhomogeneous,thomas1927calculation,fermi1927statistical}. 
The results of solving the KS equations are reported in tens of thousands of papers each year~\cite{jones2015density}. Given an approximation to the exchange-correlation (XC) energy, the KS equations are solved self-consistently. Results are limited by the quality of such approximations, and a standard problem of KS-DFT is to calculate accurate bond
dissociation curves~\cite{SWWB12}.  The difficulties are an example of strong correlation physics as electrons localize on separate nuclei~\cite{CMY08}.

Naturally, there has been considerable interest in using machine learning (ML) methods to improve DFT approximations.
Initial work~\cite{li2016understanding,snyder2012finding} focused
on the KS kinetic energy, as a sufficiently accurate approximation would allow by-passing the solving of the KS equations~\cite{brockherde2017bypassing,li2016pure}.
For XC, recent works focus on learning the XC potential (not functional) from inverse KS~\cite{jensen2018numerical}, and use it in the KS-DFT scheme~\cite{TIH96,schmidt2019machine,zhou2019toward,nagai2018neural}.
An important step forward was made last year, when it was shown that a neural network could find functionals using only three molecules, by training on both energies and densities~\cite{nagai2020completing}, 
obtaining accuracy comparable to human-designed functionals, and generalizing to yield accurate atomization energies of 148 small molecules~\cite{curtiss1991gaussian}. But this pioneering work does not yield chemical accuracy, nor approximations that work in the dissociation limit. Moreover, it uses gradient-free optimization which usually suffers from poor convergence behavior on the large number of parameters used in modern neural networks~\cite{duchi2015optimal,maheswaranathan2019guided,rios2013derivative}.

\begin{figure}[h]
\includegraphics[width=\columnwidth]{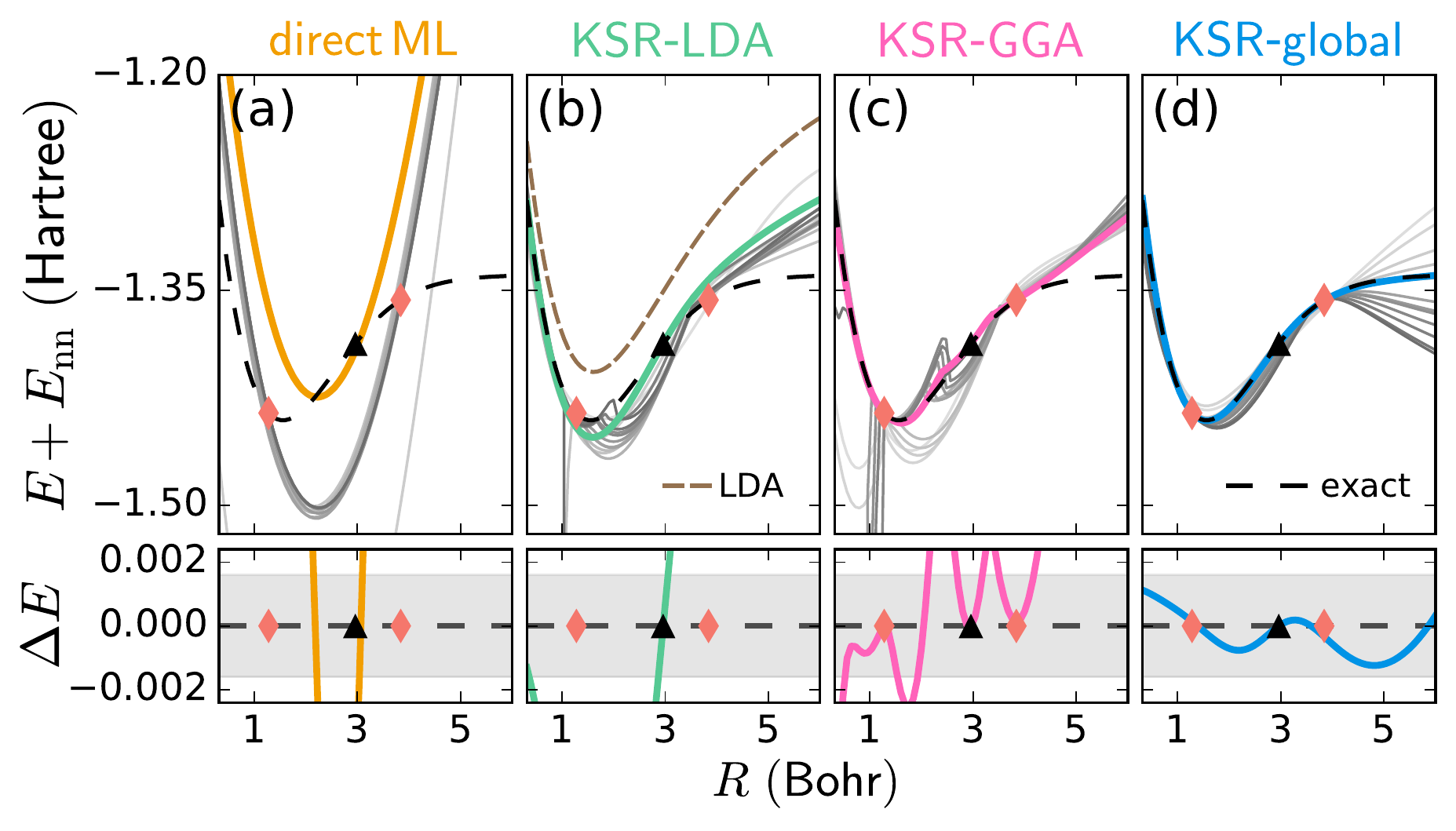}
\caption{\label{fig:distribution}
One-dimensional H$_2$ dissociation curves for several ML models trained from two molecules (red diamonds) with optimal models (highlighted in color) selected by the validation molecule at $R=3$ (black triangles). The top panel shows energy (with $E_\mathrm{nn}$, the nucleus-nucleus repulsion energy) with exact values shown by the black dashed line. The bottom panel shows difference from the exact curves with chemical accuracy in grey shadow.
(a) directly predicts $E$ from geometries and clearly fails to capture the physics from very limited data.
(b-d) shows our method (KSR) with different inputs to the model to align with the first two rungs of Jacob's ladder~\cite{perdew2005prescription} (LDA and GGA) and then global (a fully non-local functional). Uniform gas LDA~\cite{baker2015one} is shown in brown.
Grey lines denote 15 sampled functionals during training, with darker lines denoting later samples.
Atomic units used throughout.}
\end{figure}

Here, we show that all these limitations are overcome by incorporating the KS equations themselves into the neural network training by backpropagating through their iterations -- a {\it KS regularizer} (KSR) to the ML model.
In a traditional KS calculation, the XC is given, the equations are cycled to self-consistency, and all previous iterations are ignored in the final answer. In other ML work, functionals are trained on either energies alone~\cite{gilmer2017neural,schutt2017schnet,behler2007generalized,rupp2012fast}, or even densities~\cite{schmidt2019machine,zhou2019toward,moreno2020deep}, but only after convergence. By incorporating the KS equations into the training, thereby learning the relation between density and energy at every iteration, we find accurate models with very little data and much greater generalizability.

Our results are illustrated in Figure~\ref{fig:distribution}, which is for a one-dimensional mimic of H$_2$ designed for testing electronic structure methods~\cite{baker2015one}.
The distribution of curves of the ML model directly predicting E from geometries (direct ML) in (a) clearly fails to capture the physics.
Next we demonstrate KSR with neural XC functionals from the first two rungs of Jacob's ladder~\cite{perdew2005prescription}, by constraining the receptive field of the convolutional neural network~\cite{suppl}.
The local density approximation (LDA) has a receptive field of just the current point, while the generalized gradient approximation (GGA) includes the nearest neighbor points, the minimal information for computing the spatial gradient of the density. 
In (b-c), the effect of the KSR yields reasonably accurate results in the vicinity of the data, but not beyond. The KSR-LDA behaves similar to the uniform gas LDA~\cite{baker2015one}.
When an XC functional with a global receptive field is included in (d), chemical accuracy is achieved for all separations including the dissociation limit.
Similar results can be achieved for H$_4$, the one-electron self-interaction error can easily be made to vanish, and the interaction of a pair of H$_2$ molecules can be found without any training on this type of molecule (discussed below).

% KS forward
Modern DFT finds the ground-state electronic density by solving the Kohn-Sham equations:
\begin{equation}
   \bigg\{ - \frac{\nabla^2}{2} + v\s[\n](\br) \bigg\} \phi_i(\br) = \epsilon_i \phi_i(\br). 
   \label{KSeqns}
\end{equation}
The density is obtained from occupied orbitals $n(\br) = \sum_i |\phi_i (\br)|^2$.
Here $v\s[\n](\br)=v(\br)+v\H[\n](\br)+v\xc[\n](\br)$
is the KS potential consisting of the external one-body potential and the density-dependent Hartree (H) and XC potentials.
The XC potential $v\xc[\n](\br)=\delta E\xc/\delta \n(\br)$ is the functional derivative of the XC energy functional $E\xc[\n]=\int \epsilon\xc[\n](\br) \n(\br) d\br$, where $\epsilon\xc[\n](\br)$ is the XC energy per electron.
The total electronic energy $E$ is then given by the sum of the non-interacting kinetic energy $T_s[n]$, the external one-body potential energy $V[n]$, the Hartree energy $U[n]$, and XC energy $E\xc[n]$.

\begin{figure}[b]
\includegraphics[width=\columnwidth]{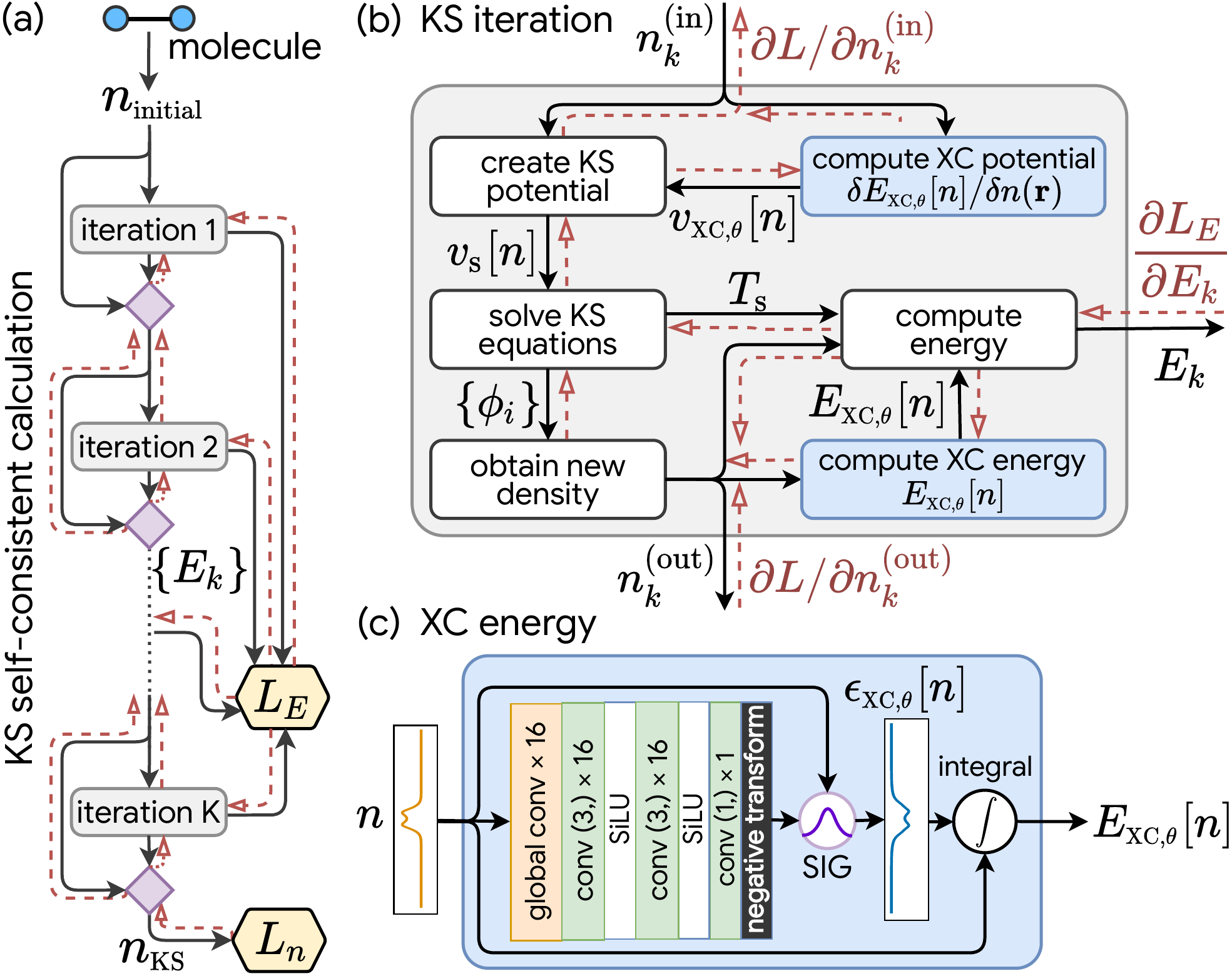}
\caption{\label{fig:ks_graph}KS-DFT as a differentiable program. Black arrows are the conventional computation flow. The gradients flow along red dashed arrows to minimize the energy loss $L_E$ and density loss $L_n$. (a) The high-level KS self-consistent calculations with linear density mixing (purple diamonds). 
(b) A single KS iteration produces $v\xctheta[n]$ and $E\xctheta[n]$ by invoking the XC energy calculation twice, once directly and once calculating a derivative using automatic differentiation.
(c) The XC energy calculation using the global XC functional.}
\end{figure}

The KS equations are in principle exact given the exact XC functional~\cite{kohn1965self, wagner2013guaranteed}, which in practice is the only term approximated in DFT.
From a computational perspective, the eigenvalue problem of Eq.~\eqref{KSeqns} is solved repeatedly until the density converges to a fixed point, starting from an initial guess.
We use linear density mixing~\cite{kresse1996efficient} to improve convergence, $n^{\sss (in)}_{k+1}=n^{\sss (in)}_{k}+\alpha(n^{\sss (out)}_{k}-n^{\sss (in)}_{k})$.
Figure~\ref{fig:ks_graph}(a) shows the unrolled computation flow.
We approximate the XC energy per electron using a neural network $\epsilon\xctheta[n]$, where $\theta$ represents the trainable parameters.
Together with the self-consistent iterations in Figure~\ref{fig:ks_graph}(b), the combined computational graph resembles a recurrent neural network~\cite{rumelhart1985learning} or deep equilibrium model~\cite{bai2019deep} with additional fixed computational components.
Density mixing improves convergence of KS self-consistent calculations and parallels the now common residual connections in deep neural networks~\cite{he2016deep} for efficient backpropagation.

% KS backward
If the neural XC functional were exact, KS self-consistent calculations would output the exact density and the intermediate energies over iterations would converge to the exact energy.
This intention can be translated into a loss function and the neural XC functional can be updated end-to-end by backpropagating through the KS self-consistent calculations. Throughout, experiments are performed in one dimension where accurate quantum solutions could be relatively easily generated via density matrix renormalization group (DMRG)~\cite{white1992density}. 
The electron-electron repulsion is $A \exp(-\kappa|x-x'|)$, and attraction to a nucleus at $x=0$ is $-A \exp(-\kappa|x|)$~\cite{suppl}.
We design the loss function as an expectation $\mathbb{E}$ over training molecules,
\begin{align}
    L(\theta)=&\underbrace{\mathbb{E}_\mathrm{train}\left[
\int dx (n\ks - n\dmrg)^2/N_e\right]}_{\mathrm{density}\, \mathrm{loss}\, L_n} \nonumber \\
&+
\underbrace{\mathbb{E}_\mathrm{train}\left[\sum_{k=1}^K 
w_k(E_k - E\dmrg)^2/N_e
\right],}_{\mathrm{energy}\, \mathrm{loss}\, L_E}
\end{align}
where $N_e$ is the number of electrons and $w_k$ are non-negative weights.
$L_n$ minimizes the difference between the final density with the exact density.
The gradient from $L_n$ backpropagates through $v\xctheta[n]$ in all KS iterations.
However, if $L_E$ only optimizes the final energy, no gradient flows through $E\xctheta[n]$ except for the final iteration. 
To make backpropagation more efficient for $E\xctheta[n]$, $L_E$ optimizes the trajectory of energies over all iterations, which directly flows gradients to early iterations~\cite{andrychowicz2016learning}.
This makes the neural XC functional output accurate $\epsilon\xc$ at each iteration, and also drives the iterations to quickly converge to the exact energy.
The optimal model is selected with minimal mean absolute energy per electron on the validation set.

% Design NN with physics intuition tailored for XC.
Hundreds of useful XC functional approximations have been proposed~\cite{mardirossian2017thirty}. Researchers typically design the symbolic form from physics intuition, with some (or no) fitting parameters.
Here we build a neural XC functional with several differentiable components with physics intuition tailored for XC in Figure~\ref{fig:ks_graph}(c).
% Global convolution
A global convolution layer captures the long range interaction, $G(n(x),\xi_p) = \frac{1}{2\xi_p}\int dx' n(x') \exp(-|x-x'|/\xi_p)$. Note two special cases retrieve known physics quantities, Hartree energy density $G(n(x),\kappa^{-1})\propto \epsilon\H$ and electronic density $G(n(x), 0)=n(x)$. Global convolution contains multiple channels and $\xi_p$ of each channel is trainable to capture interaction in different scales.
% SiLU activation and bias term in convolution
Although the rectified linear unit~\cite{nair2010rectified} is popular, we use the sigmoid linear unit (SiLU)~\cite{elfwing2018sigmoid,ramachandran2017searching} $f(x)=x/(1+\exp(-x))$ because the infinite differentiability of SiLU guarantees the smoothness of $v\xc$, the first derivative, and the second and higher order derivatives of the neural network used in the L-BFGS training~\cite{liu1989limited}.
% vanishing and negativity
We do not enforce a specific choice of $\epsilon\xc$ (sometimes called a gauge~\cite{perdew2014gedanken}), but we do enforce some conditions, primarily to aid convergence of the algorithm. We require $\epsilon\xc$ to vanish whenever the density does, and that it be negative if at all possible.
We achieved the former using the linearity of SiLU near the origin and turning off the bias terms in convolution layers. We softly impose the latter by a negative transform layer at the end, where a negative SiLU makes most output values negative.
% Self-interaction gate
Finally, we design a self-interaction gate (SIG) that mixes in a portion of $-\epsilon\H$ to cancel the self-interaction error, $\epsilon\xc^{\sss (out)}=\epsilon\xc^{\sss (in)}(1-\beta) - \epsilon\H \beta$. The portion is a gate function $\beta(N_e)=\exp(-(N_e - 1)^2 / \sigma^2)$.
When $N_e = 1$, then $\epsilon\xc^{\sss (out)}=-\epsilon\H$. For more electrons, $\sigma$ can be fixed or adjusted by the training algorithm to decide the sensitivity to $N_e$.  For H$_2$ as $R\to\infty$, $\epsilon\xc$ tends to a superposition of the negative of the Hartree energy density at each nucleus and approaches half that for H$_2^+$.

\begin{figure}[t]
\includegraphics[width=\columnwidth]{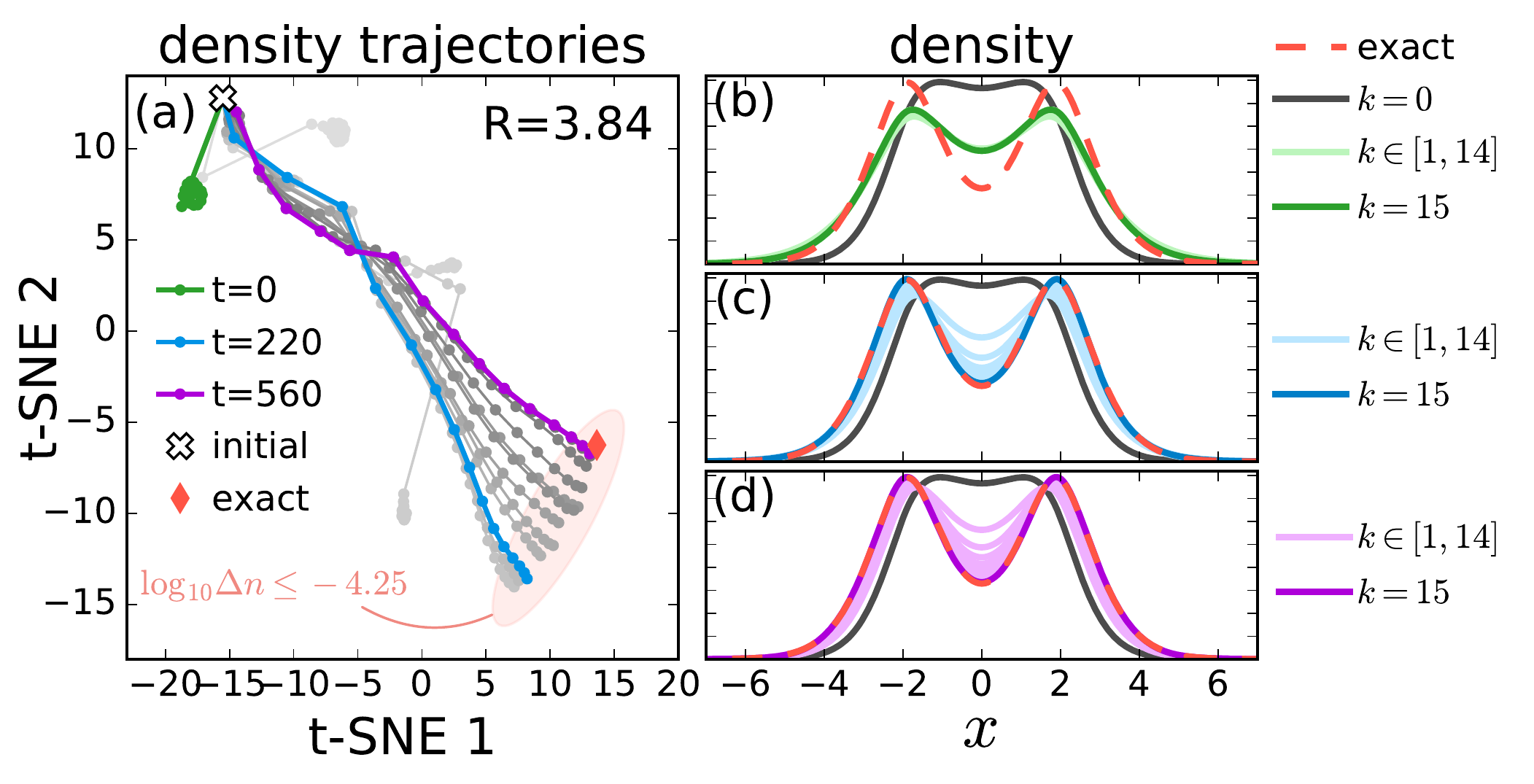}
\caption{\label{fig:density_trajectories}
(a) t-SNE visualization~\cite{maaten2008visualizing} of density trajectories (grey dots) sampled by KSR during training for $R=3.84$ from initial guess (cross) to exact density (red diamond). Darker trajectories denote later optimization steps $t$.
Note t-SNE projection does not perfectly preserve the distance between densities.
The light red ellipse illustrates the vicinity of the exact density within $\log_{10}(\int dx (n\ks - n\dmrg)^2/N_e)\leq -4.25$.
Densities from each KS iteration in trajectories are plotted in the corresponding highlighted colors for (b) $t=0$ untrained, (c) $t=220$ optimal in Figure~\ref{fig:distribution}, and (d) $t=560$ overfitting to training with bad generalization on validation.}
\end{figure}

Now we dive deeper into the outstanding generalization we observed in a simple but not easy task: predicting the entire H$_2$ dissociation curve, as shown in Figure~\ref{fig:distribution}. It is not surprising that direct ML model completely fails.
Neural networks are usually underdetermined systems as there are more parameters than training examples.
Regularization is crucial to improve generalization~\cite{Goodfellow-et-al-2016,kukavcka2017regularization}, especially when data is limited. Most existing works regularize models with particular physics prior knowledge by imposing {\it constraints} via feature engineering and preprocessing~\cite{cubuk2019screening,hollingsworth2018can}, 
architecture design~\cite{thomas2018tensor,schutt2019unifying,kondor2018covariant,seo2019differentiable} or physics-informed loss terms~\cite{raissi2019physics,sharma2018weakly}.
Another strategy is to generate extra data for training using prior knowledge: in image classification problems, data are augmented by operations like flipping and cropping given the prior knowledge that labels are invariant to those operations~\cite{krizhevsky2012imagenet}. 
KSR provides a natural data augmentation because although the exact densities and energies of only two separations are given, KSR samples different trajectories from an initial density to the exact density at each training step.
More importantly, KSR focuses on learning an XC functional that can lead the KS self-consistent calculations to converge to the exact density from the initial density.
Figure~\ref{fig:density_trajectories} visualizes the density trajectories sampled by KSR for one training separation $R=3.84$. The functional with untrained parameters ($t=0$) samples densities near the initial guess but soon learns to explore broadly and finds the trajectories toward the vicinity of the exact density.

In contrast, most existing ML functionals learn to predict the output of a single iteration from the exact density, which is a poor surrogate for the full self-consistent calculations~\cite{tucker2017rebar}. These standard ML models have two major shortcomings. First, the exact density is unknown for new systems, so the model is not expected to behave correctly on unseen initial densities for KS calculations. Second, even if a model is trained on many densities for single iteration prediction, it is not guaranteed to converge the self-consistent calculations to a good solution~\cite{ross2010efficient}. On the other hand, since KSR allows the model access to all the KS iterations, it learns to optimize the entire self-consistent procedure to avoid the error accumulation from greedy optimization of single iterations.
Further comparison for training neural XC functionals without or with ``weaker'' KSR is in the supplemental material~\cite{suppl}.

\begin{figure}[t]
\includegraphics[width=\columnwidth]{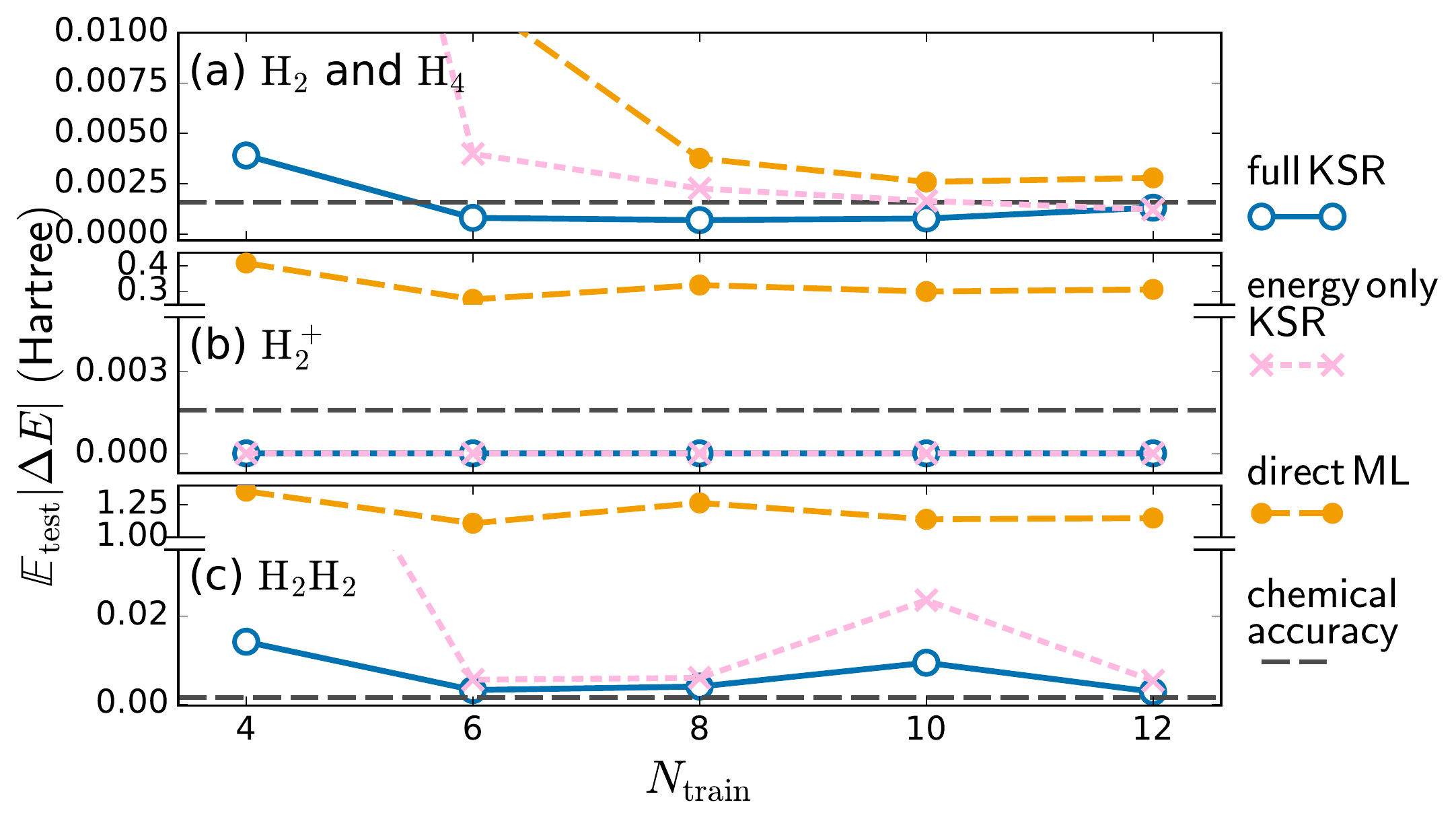}
\caption{\label{fig:generalization}
Test generalization of models as a function of the total number of training examples $N_\mathrm{train}$: full KSR (blue), energy only KSR (pink) and direct ML (orange) on (a) holdout H$_2$ and H$_4$, and unseen types of molecules (b) H$_2^+$ (c) H$_2$H$_2$.
Black dashed lines show chemical accuracy. See the supplemental material~\cite{suppl} for training details.
}
\end{figure}

Next we retrain our neural XC functional with KSR on $N_\mathrm{train}/2$ examples each of H$_2$ and H$_4$ molecules.
Figure~\ref{fig:generalization} shows the prediction accuracy of KSR with both energy and density loss (full KSR), in comparison to KSR with only energy loss (energy only KSR) and direct ML model.
We compute the energy mean absolute error on the holdout sets of H$_2$ ($R\in[0.4,6]$) and H$_4$ ($R\in[1.04,6]$).
The average mean absolute error of H$_2$ and H$_4$ with various $N_\mathrm{train}$ is shown in Figure~\ref{fig:generalization}(a).
Full KSR has the lowest error at minimum $N_\mathrm{train}=4$, reaching chemical accuracy at $6$.
As the size of the training set increases, energy only KSR reaches chemical accuracy at $N_\mathrm{train}=10$, but direct ML model never does (even at $20$).
Then we test models on unseen types of molecules.
In Figure~\ref{fig:generalization}(b), both KSR models have perfect prediction on H$_2^+$ ($R\in[0.64,8.48]$) because of the self-interaction gate in the neural XC functionals, while direct ML models always have large errors.
Finally we take a pair of equilibrium H$_2$ and separate them with $R=0.16$ to $9.76$ Bohr, denoted as H$_2$H$_2$.
KSR models generalize much better than ML for ``zero-shot'' prediction~\cite{mirhoseini2020chip}, where H$_2$H$_2$ has never been exposed to the model during training.

\begin{figure}[t]
\includegraphics[width=\columnwidth]{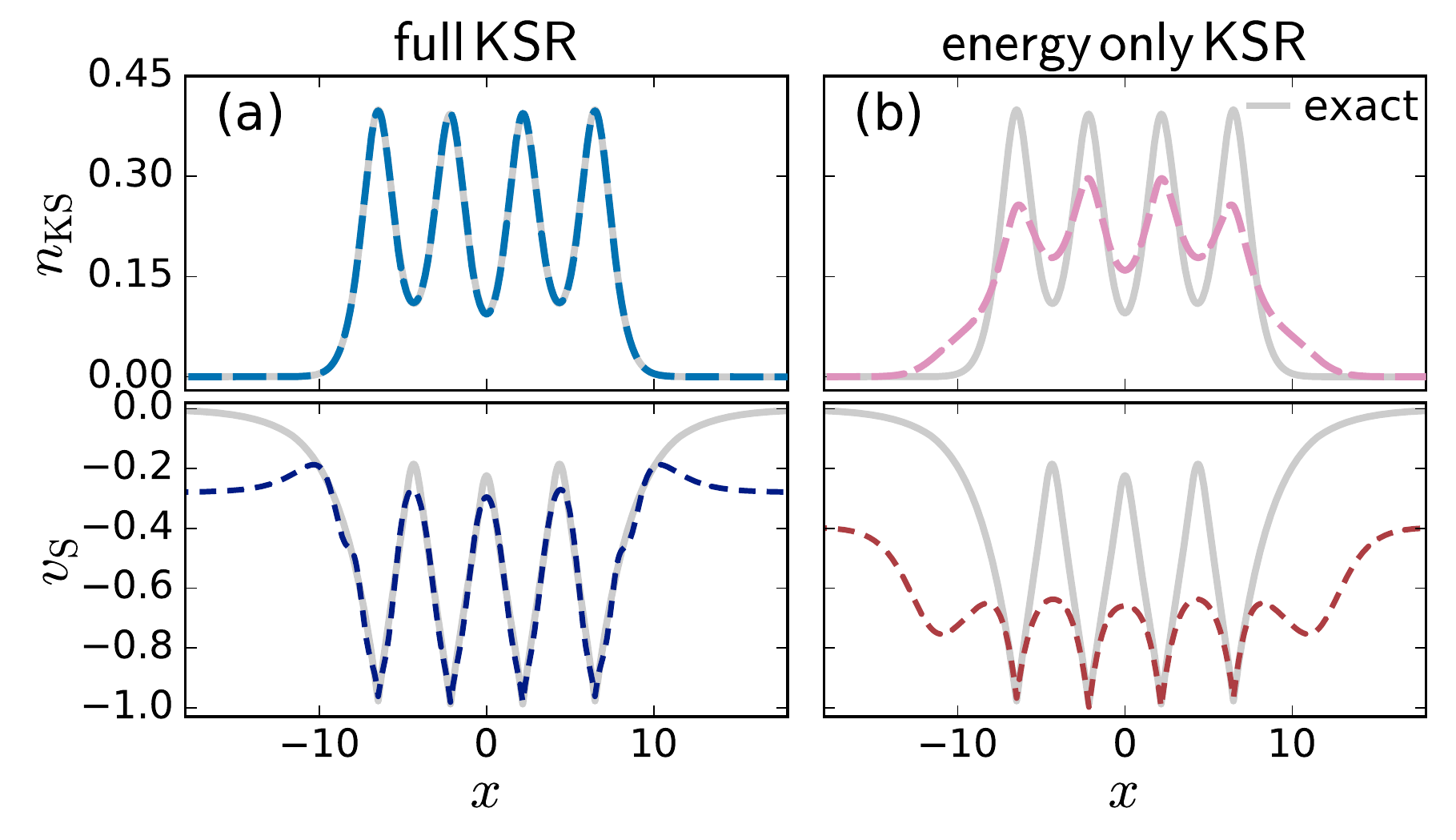}
\caption{\label{fig:h4_density}
Density and KS potential of H$_4$ with $R=4.32$ from neural XC functionals trained with (a) full KSR (blue) and (b) energy only KSR (pink) on training set of size $N_\mathrm{train}=20$. Exact curves are in grey. $v\s$ are shifted by a constant for better comparison.
}
\end{figure}

Why is the density important in training, and what use are the non-converged iterations?
The density is the functional derivative of the energy with respect to the potential, so it gives the exact slope of the energy with respect to any change in the potential, including stretching (or compressing) the bond.
Thus the density implicitly contains energetic information including the correct derivative at that point in the binding curve.
KS iterations produce information about the functional in the vicinity of the minimum.  
During training, the network learns to construct a functional with both the correct minimum and all correct derivatives at this minimum.
In the paradigm of differentiable programming, density is the hidden state carrying the information through the recurrent structure in Figure~\ref{fig:ks_graph}(a).
Correct supervision from $L_n$ greatly helps generalization from very limited data, see $N_\mathrm{train}\leq 6$ in Figure~\ref{fig:generalization}.
But as $N_\mathrm{train}$ increases, both KSR with and without $L_n$ perform well in energy prediction.
We show the solution of H$_4$ with $R=4.32$ in Figure~\ref{fig:h4_density}.
With $L_n$, the density is clearly much accurate than KSR without $L_n$ ($\int (n\ks-n\dmrg)^2 dx=9.2\times10^{-5}$ versus $9.8\times10^{-2}$).
Then we compute the corresponding exact $v\s$ using inverse KS method~\cite{jensen2018numerical}.
Both functionals do not reproduce the exact $v\s$.
However, functional trained with $L_n$ recovered most of the KS potential.
Unlike previous works~\cite{schmidt2019machine,zhou2019toward,nagai2018neural} that explicitly included the KS or XC potential into the loss function, our model never uses the exact KS potential.
In our KSR setup, the model aims at predicting $\epsilon\xc$, from which the derived $v\s$ yields accurate density.
Therefore, predicting $v\xc$ is a side product.
We also address some concerns on training explicitly with $v\xc$.
One artifact is that generating the exact $v\s$ requires an additional inverse calculation, which is known to be numerically unstable~\cite{jensen2018numerical}.
\citet{schmidt2019machine} observe outliers while generating training $v\xc$ from inverse KS.
While $v\xc$ is a fascinating and useful object for theoretical study, because its relation to the density is extremely delicate, it is far more practical to simply use the density to train on~\cite{nagai2020completing}.

Differentiable programming blurs the boundary between physics computation and ML.
Our results for KS-DFT serve as proof of principle for rethinking computational physics in this new paradigm.
Although there is no explicit limitation of our algorithm to one dimension, we expect practical challenges with real molecules, which will require rewriting or extending a mature DFT code to support automatic differentiation.
For example, our differentiable eigensolver for dense matrices~\cite{mat_cookbook} is not suitable for large problems, and will need to be replaced with methods for partial eigendecomposition of sparse matrices~\cite{Lee2007-bo,xie2020automatic}.
Beyond density functionals, in principle all heuristics in DFT calculations, e.g., initial guess, density update, preconditioning, basis sets, even the entire self-consistent calculations as a meta-optimization problem~\cite{andrychowicz2016learning}, could be learned and optimized while maintaining rigorous physics -- getting the best of both worlds.

\begin{acknowledgments}
The authors thank Michael Brenner, Sam Schoenholz, Lucas Wagner, and Hanjun Dai for helpful discussion. K.B. supported by NSF CHE-1856165, R.P. by DOE DE-SC0008696.
Code is available at \url{https://github.com/google-research/google-research/tree/master/jax_dft}
\end{acknowledgments}

\bibliography{references}

\end{document}

\begin{CJK*}{UTF8}{gbsn}

\title{Supplemental Material\\Building prior knowledge into machine-learned physics: \\Kohn-Sham equations as a regularizer}% Force line breaks with \\

\author{Li Li (李力)$^{1}$}
\email[email: ]{leeley@google.com}
\author{Stephan Hoyer$^{1}$}
\author{Ryan Pederson$^{2}$}
\author{Ruoxi Sun (孙若溪)$^{1}$}
\author{Ekin D.~Cubuk$^{1}$}
\author{Patrick Riley$^{1}$}
\author{Kieron Burke$^{2,3}$}
\affiliation{
$^1$ Google Research, Mountain View, CA 94043, USA \\
$^2$ Department of Physics and Astronomy, University of California, Irvine, CA 92697, USA \\
$^3$ Department of Chemistry, University of California, Irvine, CA 92697, USA}

\let\clearpage\relax
\maketitle
\tableofcontents
\end{CJK*}

\section{1D Model systems}
In 1D, we utilize exponential Coulomb interactions to mimic the standard 3D Coulomb potential, 
\begin{equation}
    v_{\exp}(x)=A \exp(-\kappa|x|),
\end{equation}
where $A=1.071295$ and $\kappa^{-1}=2.385345$~\cite{baker2015one}. Within this model, the external one-body potential for a nuclei of atomic number $Z$ and position $x'$ is represented by $-Z \, v_{\exp}(x - x')$. The external potential for arbitrary molecular systems and geometries is modeled as
\begin{equation}\label{eq:external_potential}
    v(x) = -\sum_j Z_j v_{\exp}(x - x_j) \, .
\end{equation}
For example, a 1D H$_2$ molecule at separation $R = 4$ can be represented by $v(x) = -v_{\exp}(x - 2) - v_{\exp}(x + 2)$. The repulsion between electrons at positions $x$ and $x'$ is given by the two-body potential $v\ee(x-x') = v_{\exp}(x - x')$. We represent all systems on a 1D grid of~$m = 513$ points each separated by a distance $h = 0.08$. The center grid point is at the origin, $x = 0$, and the range of grid points is $x \in \{-20.48, \dots, 20.48\}$.  For consistency, all nuclei positions reside on grid points in calculations. In this convention, all molecules in this work are either symmetric about the origin $x = 0$ or $x = 0.04$, depending on the separation between nuclei. 

\section{DMRG calculation details}
The real-space interacting Hamiltonian for a 1D system of lattice spacing $h$ becomes in second quantized notation,
\begin{equation}
    H = \frac{5}{4h^2} \sum_{j,\sigma} n_{j\sigma} - \frac{2}{3h^2} \sum_{\langle {i, j} \rangle,\sigma}  c^\dagger_{i\sigma}c_{j\sigma} + \frac{1}{24h^2} \sum_{\langle \langle {i, j} \rangle \rangle,\sigma} c^\dagger_{i\sigma}c_{j\sigma} + \sum_j v(x_j) \, n_j + \sum_{ij} v\ee(x_i-x_j)\,  n_i \, n_j \, ,
\end{equation}
where the operator $c^\dagger_{j\sigma}$ creates (and $c_{j\sigma}$ annihilates) an electron of spin $\sigma$ on site $j$, $n_{j \sigma} = c^\dagger_{j\sigma} c_{j\sigma}$, and $n_j = n_{j \uparrow} + n_{j \downarrow}$. The single and double brackets below the sums indicate sums over nearest and next nearest neighbors, respectively. The hopping term coefficients are determined by the $4$-th order central finite difference approximation to the second derivative. The Hamiltonian is solved using DMRG to obtain highly accurate ground-state energies and densities. Calculations are performed using the ITensor library~\cite{itensor} with an energy convergence threshold of $10^{-7}$ Ha.

\section{KS calculation details}
\subsection{Local Density Approximation}
In our 1D model the electron repulsion is an exponential interaction. To implement a local density approximation (LDA) for this interaction we use Ref.~\cite{baker2015one} which provides the exponentially repelling uniform gas exchange energy analytically and an accurate parameterized model for the correlation energy. We use this specific implementation for all LDA calculations.        

\subsection{Initial density}
We solve the Schr{\"o}dinger equation of the non-interacting system with the external potential $v(x)$ defined in Eq.~\ref{eq:external_potential},
\begin{align}
   \bigg\{ - \frac{\nabla^2}{2} + v(x) \bigg\} \phi_i(x) = \epsilon_i \, \phi_i(x). 
   \label{eq:non_interacting_eqns}
\end{align}
The density is the square sum of all the occupied orbitals $n(x) = \sum_i |\phi_i (x)|^2$. In all the KS self-consistent calculations presented in this work, we use the density of the non-interacting system with external potential $v(x)$ as the initial density.

\subsection{Linear density mixing}
Linear density mixing is a well-known strategy to improve the convergence of the KS self-consistent calculation,
\begin{equation}
  n^{\sss (in)}_{k+1}=n^{\sss (in)}_{k}+\alpha(n^{\sss (out)}_{k}-n^{\sss (in)}_{k}).
\end{equation}
In this work, we apply an exponential decay on the mixing factor $\alpha=0.5\times0.9^{k-1}$.

\subsection{XC potential from automatic differentiation}
The XC functional $\epsilon\xc[n]:\mathbb{R}^m\to\mathbb{R}^m$ is a mapping from density $n\in\mathbb{R}^m$ to XC energy density $\epsilon\xc\in\mathbb{R}^m$, where $m$ is the number of grids. Instead of hand-deriving the functional derivative, we use automatic differentiation in JAX~\cite{jax2018github} to compute 
\begin{equation}
    v\xc=\frac{\delta E\xc}{\delta n}=\frac{\delta\int n\,\epsilon\xc[n]dx}{\delta n}.
\end{equation}

The XC energy density $\epsilon\xc[n]$ is denoted as \verb!xc_energy_density_fn!. It takes the \verb!density!, a float array with size $m$ as input argument and returns a float array with size $m$. This function can be a conventional physics XC functional (e.g., LDA) or a neural XC functional.

The XC energy $E\xc$ can be computed by the following function,
\begin{lstlisting}[language=Python]
def get_xc_energy(density, xc_energy_density_fn, grids):
  """Gets the xc energy by discretizing the following integral.

  E_xc = \int density * xc_energy_density_fn(density) dx.

  Args:
    density: Float numpy array with shape (num_grids,).
    xc_energy_density_fn: function takes density and returns float numpy array
        with shape (num_grids,).
    grids: Float numpy array with shape (num_grids,).

  Returns:
    Float.
  """
  return jnp.dot(xc_energy_density_fn(density), density) * utils.get_dx(grids)
\end{lstlisting}

Then the functional derivative $v\xc=\delta E\xc/\delta n$ can be computed via \verb!jax.grad! function,
\begin{lstlisting}[language=Python]
def get_xc_potential(density, xc_energy_density_fn, grids):
  """Gets xc potential.

  The xc potential is derived from xc_energy_density through automatic
  differentiation.

  Args:
    density: Float numpy array with shape (num_grids,).
    xc_energy_density_fn: function takes density and returns float numpy array
        with shape (num_grids,).
    grids: Float numpy array with shape (num_grids,).

  Returns:
    Float numpy array with shape (num_grids,).
  """
  return jax.grad(get_xc_energy)(
      density, xc_energy_density_fn, grids) / utils.get_dx(grids)
\end{lstlisting}

The \verb!jax.grad! function computes the gradient of \verb!get_xc_energy! function with respect to the first argument \verb!density! using automatic differentiation. Both functions can be found in the \verb!scf! module in the JAX-DFT library.

\subsection{Symmetry}

The training molecules used in this paper are symmetric to their centers.
We define the symmetry operation on functions on the grids $\mathcal{S}:\mathbb{R}^m\to\mathbb{R}^m$. It flips a function at the center and averages with itself. In each KS iteration, we enforce the symmetry on the XC functional, $\epsilon\xc[n] \to \mathcal{S}\big(\epsilon\xc[\mathcal{S}(n)]\big)$. So $E\xc$ and $v\xc$ are transformed as
\begin{align}
E\xc &= \int n\,\epsilon\xc[n]dx \to \int  n\,\mathcal{S}\big(\epsilon\xc[\mathcal{S}(n)]\big)dx \\
v\xc &= \frac{\delta\int n\,\epsilon\xc[n]dx}{\delta n} \to \frac{\delta\int  n\,\mathcal{S}\big(\epsilon\xc[\mathcal{S}(n)]\big)dx}{\delta n}.
\end{align}
Before the output of each KS iteration, $n_k^{\sss (out)}\to \mathcal{S}(n_k^{\sss (out)})$.

We note that applying the symmetry restriction does not change the model performance of the molecules around equilibrium. However, because both stretched H$_2$ and H$_4$ have vanishing KS gaps, the KS self-consistent calculation is difficult to converge. The gradient information for the KSR relies on the stable output of the KS self-consistent calculation. In realistic 3D DFT codes, enforcing symmetry is one common strategy to improve the convergence of KS self-consistent calculations. Therefore, we applied symmetry restriction so that KSR can obtain stable gradient information in the stretched limit cases.

\section{Training, validation and test}

\subsection{Weights in trajectory loss}
We use $w_k=0.9^{K-k}H(k-10)$, where $H$ is the Heaviside function and $K$ is the total number of iterations.

\subsection{Number of KS iterations}

We first run KS calculations with the standard uniform gas LDA functional~\cite{baker2015one}. The number of iterations to converge the largest separation for different molecules are around 8, 25, 5, 6 for H$_2$ , H$_4$, H$_2^+$, H$_2$H$_2$ respectively. During the training of neural XC functionals with KSR, we use a fixed number of iterations that is greater than or equal to the estimation from LDA for each type of molecules so it is sufficient for convergence.
The number of iterations for different molecules are listed in Table~\ref{table:iterations}.

\begin{table}[htb]
\caption{\label{table:iterations} Number of iterations for different molecules.}
\begin{minipage}{0.5\textwidth}
\begin{center}
\begin{ruledtabular}
\begin{tabular}{lcccc}
& H$_2$ & H$_4$ & H$_2^+$ & H$_2$H$_2$\\
\hline
train & 15 & 40 & -- & --\\
validation & 15 & 40 & -- & --\\
test & 15 & 40 & 5 & 10\\
\end{tabular}
\end{ruledtabular}
\end{center}
\end{minipage}
\end{table}

\subsection{Dataset for learning H$_2$ dissociation from two molecules}
H$_2$ dissociation curves in Figure~1 are trained from exact densities and energies of two molecules. One is a compressed H$_2$ ($R=1.28$) and the other is a stretched H$_2$ ($R=3.84$) molecule. The optimal checkpoint is selected by a validation molecule with $R=2.96$.

\subsection{Dataset for learning and predicting several types of molecules}

\subsubsection{Training molecules}

The distances between nearby atoms for training molecules used in Figure~4 are listed in Table~\ref{table:training_molecules}.

\begin{table}[htb]
\caption{\label{table:training_molecules} Distances between nearby atoms for training molecules used in Figure~4.}
\begin{ruledtabular}
\begin{tabular}{c|lllllllllll|llllllllll}
$N_\mathrm{train}$ & & & & & H$_2$ & & & & & & & & & & & & H$_4$ & & & &\\
\hline
4  &        &        & $1.28$ &        &        &        &        & $3.84$ &        &        & &        &        & $2.08$ &        &        & $3.36$ &        &        &        &       \\

6  & $0.48$ &        & $1.28$ &        &        &        &        & $3.84$ &        &        & & $1.28$ &        & $2.08$ &        &        & $3.36$ &        &        &        &       \\

8  & $0.48$ &        & $1.28$ &        &        & $3.04$ &        & $3.84$ &        &        & & $1.28$ &        & $2.08$ &        &        & $3.36$ &        &        & $4.48$ &       \\

10 & $0.48$ &        & $1.28$ &        &        & $3.04$ &        & $3.84$ &        & $4.64$ & & $1.28$ &        & $2.08$ &        &        & $3.36$ &        & $4.00$ & $4.48$ &       \\

12 & $0.48$ &        & $1.28$ &        & $2.40$ & $3.04$ &        & $3.84$ &        & $4.64$ & & $1.28$ &        & $2.08$ &        & $3.04$ & $3.36$ &        & $4.00$ & $4.48$ &       \\

14 & $0.48$ &        & $1.28$ &        & $2.40$ & $3.04$ & $3.52$ & $3.84$ &        & $4.64$ & & $1.28$ &        & $2.08$ &        & $3.04$ & $3.36$ & $3.68$ & $4.00$ & $4.48$ &       \\

16 & $0.48$ &        & $1.28$ & $1.76$ & $2.40$ & $3.04$ & $3.52$ & $3.84$ &        & $4.64$ & & $1.28$ &        & $2.08$ & $2.56$ & $3.04$ & $3.36$ & $3.68$ & $4.00$ & $4.48$ &       \\

18 & $0.48$ &        & $1.28$ & $1.76$ & $2.40$ & $3.04$ & $3.52$ & $3.84$ & $4.16$ & $4.64$ & & $1.28$ &        & $2.08$ & $2.56$ & $3.04$ & $3.36$ & $3.68$ & $4.00$ & $4.48$ & $4.80$\\

20 & $0.48$ & $0.80$ & $1.28$ & $1.76$ & $2.40$ & $3.04$ & $3.52$ & $3.84$ & $4.16$ & $4.64$ & & $1.28$ & $1.76$ & $2.08$ & $2.56$ & $3.04$ & $3.36$ & $3.68$ & $4.00$ & $4.48$ & $4.80$\\
% \hline
% 4 & $1.28,3.84$ & $2.08,3.36$\\
% 6 & $0.48,1.28,3.84$ & $1.28,2.08,3.36$\\
% 8 & $0.48,1.28,3.04,3.84$ & $1.28,2.08,3.36,4.48$\\
% 10 & $0.48,1.28,3.04,3.84,4.64$ & $1.28,2.08,3.36,4.00,4.48$\\
% 12 & $0.48,1.28,2.40,3.04,3.84,4.64$ & $1.28,2.08,3.04,3.36,4.00,4.48$\\
% 14 & $0.48,1.28,2.40,3.04,3.52,3.84,4.64$ & $1.28,2.08,3.04,3.36,3.68,4.00,4.48$\\
% 16 & $0.48,1.28,1.76,2.40,3.04,3.52,3.84,4.64$ & $1.28,2.08,2.56,3.04,3.36,3.68,4.00,4.48$\\
% 18 & $0.48,1.28,1.76,2.40,3.04,3.52,3.84,4.16,4.64$ & $1.28,2.08,2.56,3.04,3.36,3.68,4.00,4.48,4.80$\\
% 20 & $0.48,0.80,1.28,1.76,2.40,3.04,3.52,3.84,4.16,4.64$ & $1.28,1.76,2.08,2.56,3.04,3.36,3.68,4.00,4.48,4.80$\\
\end{tabular}
\end{ruledtabular}
\end{table}

\subsubsection{Validation molecules}
The distances between nearby atoms for validation molecules used in Figure~4 are listed in Table~\ref{table:validation_molecules}.

\begin{table}[htb]
\caption{\label{table:validation_molecules} The distances between nearby atoms for validation molecules used in Figure~4. The validation set is fixed for calculations with $4\leq N_\mathrm{train}\leq 20$.}
\begin{minipage}{0.3\textwidth}
\begin{center}
\begin{ruledtabular}
\begin{tabular}{cl}
Molecule & $R$ \\
\hline
H$_2$ & $1.68,2.96,4.40,5.52$\\
H$_4$ & $1.84,2.64,3.28,5.04$\\
\end{tabular}
\end{ruledtabular}
\end{center}
\end{minipage}
\end{table}

\subsubsection{Test molecules}

The distances between nearby atoms for test molecules used in Figure~4 are listed in Table~\ref{table:test_molecules}.

\begin{table}[htb]
\caption{\label{table:test_molecules} The distances between nearby atoms for test molecules used in Figure~4. The test set is fixed for calculations with $4\leq N_\mathrm{train}\leq 20$.}
\begin{ruledtabular}
\begin{tabular}{cl}
Molecule & $R$ \\
\hline
\multirow{2}{*}{H$_2$} & $0.40, 0.56, 0.72, 0.88, 1.04, 1.20, 1.36, 1.52, 1.84, 2.00, 2.16, 2.32, 2.48, 2.64, 2.80, 3.12, 3.28, 3.44, 3.60, 3.76, $ \\
 & $ 3.92, 4.08, 4.24,4.56, 4.72, 4.88, 5.04, 5.20, 5.36, 5.68, 5.84, 6.00$\\
\hline
\multirow{2}{*}{H$_4$} & $1.04, 1.20, 1.36, 1.52, 1.68, 2.00, 2.16, 2.32, 2.48, 2.80, 2.96, 3.12, 3.44, 3.60, 3.76, 3.92, 4.08, 4.24, 4.40, 4.56$\\
 & $4.72, 4.88, 5.20, 5.36, 5.52, 5.68, 5.84, 6.00$ \\
\hline
\multirow{3}{*}{H$_2^+$} & $0.64, 0.80, 0.96, 1.12, 1.28, 1.44, 1.60, 1.76, 1.92, 2.08, 2.24, 2.40, 2.48, 2.56, 2.64, 2.72, 2.88, 3.04, 3.20, 3.36, $\\
& $3.52, 3.68, 3.84, 4.00, 4.16, 4.32, 4.48, 4.64, 4.80, 4.96, 5.12, 5.28, 5.44, 5.60, 5.76, 5.92, 6.08, 6.24, 6.40, 6.56, $ \\
& $6.72, 6.88, 7.04, 7.20, 7.36, 7.52, 7.68, 7.84, 8.00, 8.16, 8.32, 8.48$ \\
\hline
\multirow{2}{*}{H$_2$H$_2$} & $0.16, 0.48, 0.80, 1.12, 1.44, 1.76, 2.08, 2.40, 2.72, 3.04, 3.36, 3.68, 4.00, 4.32, 4.64, 4.96, 5.28, 5.60, 5.92, 6.24, $\\
& $6.56, 6.88, 7.20, 7.52, 7.84, 8.16, 8.48, 8.80, 9.12, 9.44, 9.76$ \\
\end{tabular}
\end{ruledtabular}
\end{table}

\subsubsection{Test errors}
\begin{figure}[htb]
\includegraphics[width=0.5\columnwidth]{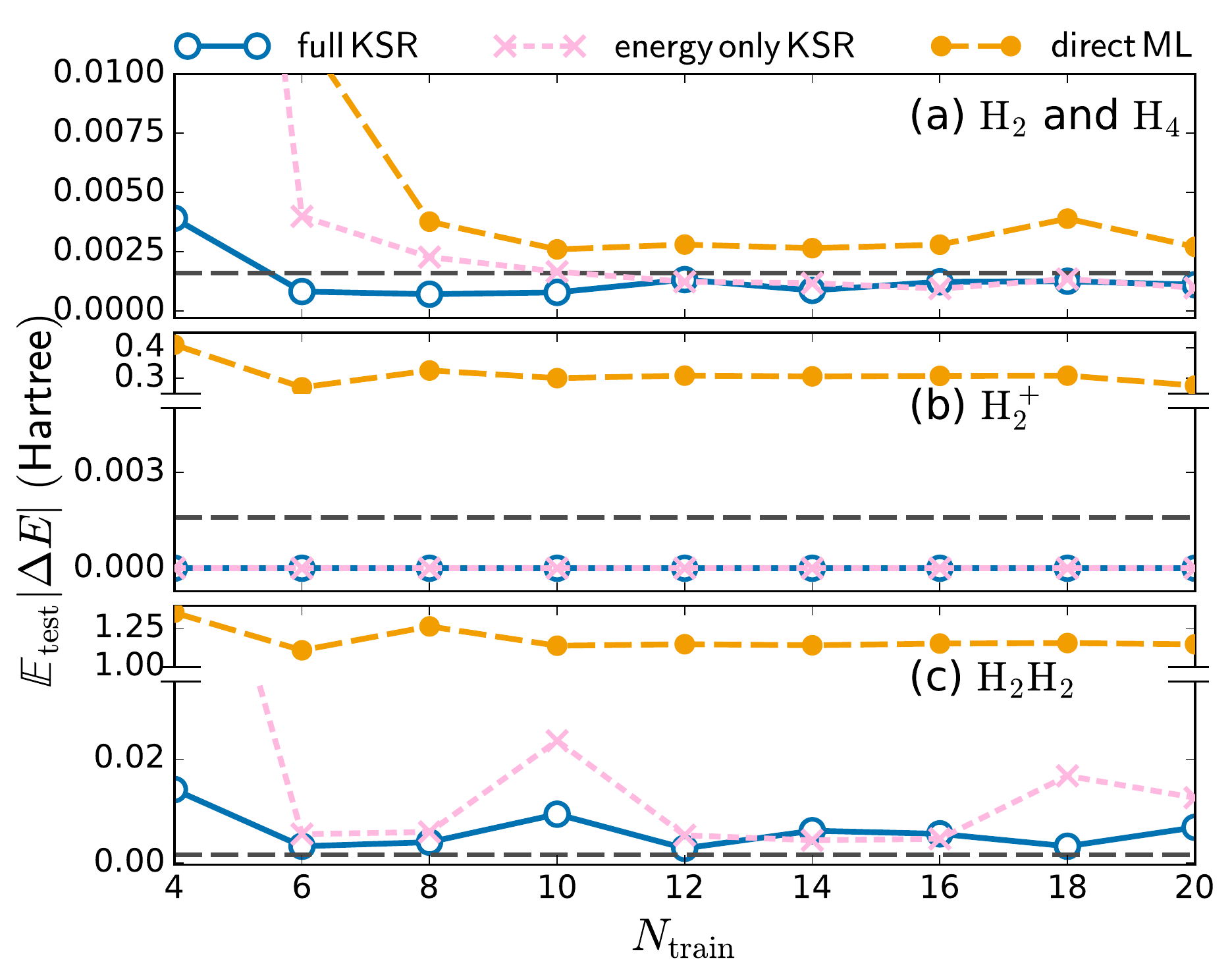}
\caption{\label{fig:generalization_to_20}
Test generalization of ML models as a function of the total number of training examples $N_\mathrm{train}$: full KSR (blue), energy only KSR (pink) and direct ML (orange) on (a) holdout H$_2$ and H$_4$, and unseen types of molecules (b) H$_2^+$ (c) H$_2$H$_2$.
Black dashed lines show chemical accuracy. All the numerical values are listed in Table~\ref{table:generalization}.
}
\end{figure}

\begin{table}[htb]
\caption{\label{table:generalization}Numerical values of test errors plotted in Figure~\ref{fig:generalization_to_20}.}
%\begin{ruledtabular}
\resizebox{0.95\columnwidth}{!}{
\begin{tabular}{ccccccccccc}
\hline
\hline
\multirow{2}{*}{Model} & \multirow{2}{*}{Molecule} & & & & & $N_\mathrm{train}$ & & & & \\
& & 4 & 6 & 8 & 10 & 12 & 14 & 16 & 18 & 20 \\
\hline
\multirow{3}{*}{KSR $L_n + L_E$} & H$_2$ and H$_4$ & $3.91\times 10^{-3}$ & $8.15\times 10^{-4}$ & $7.04\times 10^{-4}$ & $7.82\times 10^{-4}$ & $1.32\times 10^{-3}$ & $8.64\times 10^{-4}$ & $1.23\times 10^{-3}$ & $1.25\times 10^{-3}$ & $1.11\times 10^{-3}$ \\
& H$_2^+$ & $1.71\times 10^{-5}$ & $1.71\times 10^{-5}$ & $1.71\times 10^{-5}$ & $1.71\times 10^{-5}$ & $1.71\times 10^{-5}$ & $1.71\times 10^{-5}$ & $1.71\times 10^{-5}$ & $1.71\times 10^{-5}$ & $1.71\times 10^{-5}$ \\
& H$_2$H$_2$ & $1.42\times 10^{-2}$ & $3.23\times 10^{-3}$ & $4.02\times 10^{-3}$ & $9.41\times 10^{-3}$ & $2.83\times 10^{-3}$ & $6.23\times 10^{-3}$ & $5.62\times 10^{-3}$ & $3.23\times 10^{-3}$ & $6.87\times 10^{-3}$ \\
\hline
\multirow{3}{*}{KSR $L_E$} & H$_2$ and H$_4$ & $4.82\times 10^{-2}$ & $3.99\times 10^{-3}$ & $2.27\times 10^{-3}$ & $1.66\times 10^{-3}$ & $1.23\times 10^{-3}$ & $1.17\times 10^{-3}$ & $9.37\times 10^{-4}$ & $1.35\times 10^{-3}$ & $9.73\times 10^{-4}$ \\
& H$_2^+$ & $1.71\times 10^{-5}$ & $1.71\times 10^{-5}$ & $1.71\times 10^{-5}$ & $1.71\times 10^{-5}$ & $1.71\times 10^{-5}$ & $1.71\times 10^{-5}$ & $1.71\times 10^{-5}$ & $1.71\times 10^{-5}$ & $1.71\times 10^{-5}$ \\
& H$_2$H$_2$ & $9.19\times 10^{-2}$ & $5.56\times 10^{-3}$ & $5.99\times 10^{-3}$ & $2.36\times 10^{-2}$ & $5.35\times 10^{-3}$ & $4.38\times 10^{-3}$ & $4.67\times 10^{-3}$ & $1.68\times 10^{-2}$ & $1.26\times 10^{-2}$ \\
\hline
\multirow{3}{*}{ML} & H$_2$ and H$_4$ & $4.95\times 10^{-2}$ & $1.18\times 10^{-2}$ & $3.77\times 10^{-3}$ & $2.60\times 10^{-3}$ & $2.80\times 10^{-3}$ & $2.65\times 10^{-3}$ & $2.79\times 10^{-3}$ & $3.89\times 10^{-3}$ & $2.70\times 10^{-3}$ \\
& H$_2^+$ & $4.10\times 10^{-1}$ & $2.71\times 10^{-1}$ & $3.26\times 10^{-1}$ & $3.01\times 10^{-1}$ & $3.09\times 10^{-1}$ & $3.07\times 10^{-1}$ & $3.08\times 10^{-1}$ & $3.09\times 10^{-1}$ & $2.76\times 10^{-1}$
 \\
& H$_2$H$_2$ & $1.36$ & $1.11$ & $1.27$ & $1.14$ & $1.15$ & $1.14$ & $1.15$ & $1.16$ & $1.15$ \\
\hline
\hline
\end{tabular}}
%\end{ruledtabular}

\end{table}

We extend the plot of test errors in Figure~4 to $N_\mathrm{train}=20$ in Figure~\ref{fig:generalization_to_20} and list all the numerical values in Table~\ref{table:generalization}.

\subsubsection{Dissociation curve of H$_2$, H$_4$, H$_2^+$ and H$_2$H$_2$}

\begin{figure}[htb]
\includegraphics[width=0.5\columnwidth]{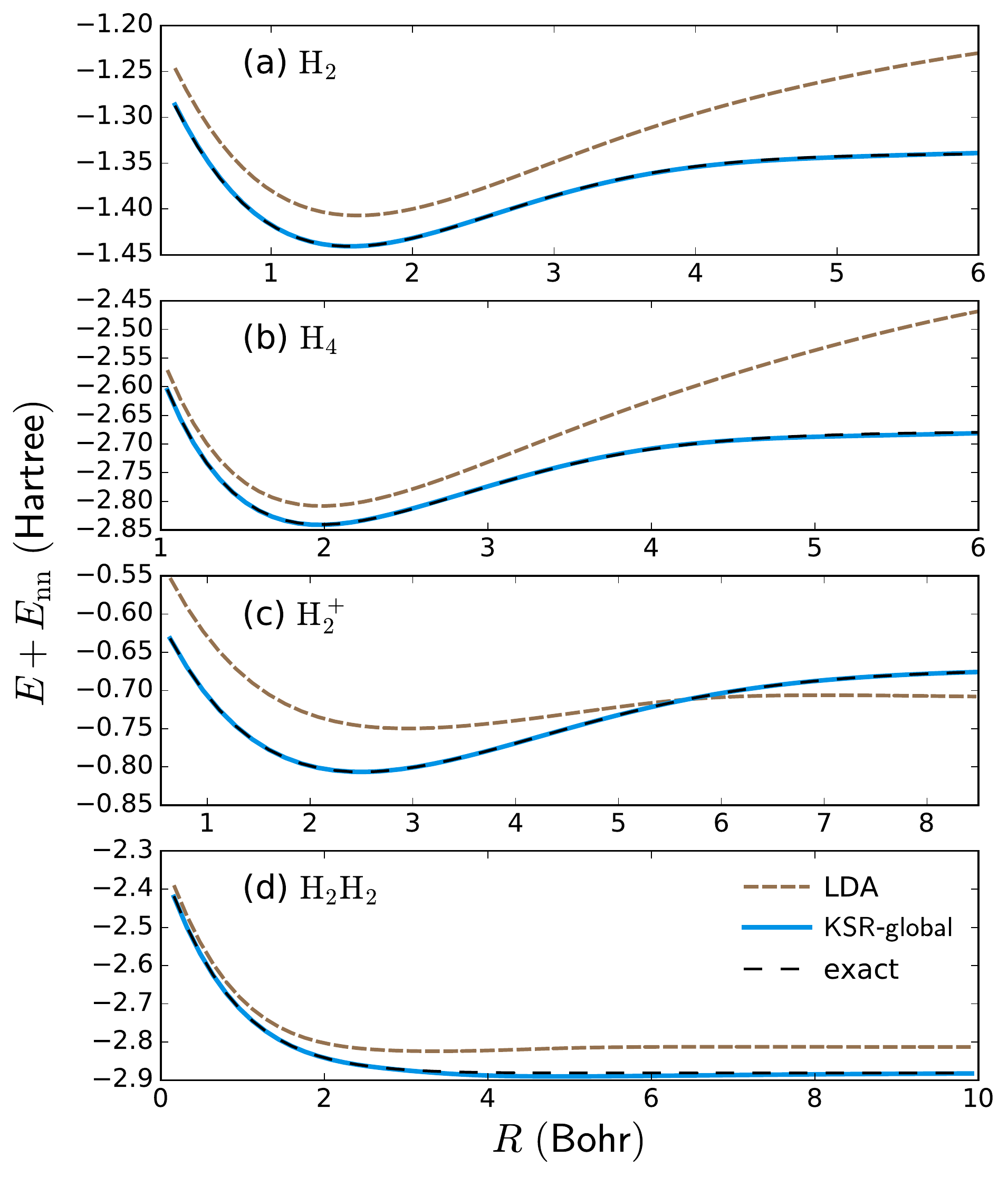}
\caption{\label{fig:pes}
Dissociation curves of (a) H$_2$, (b) H$_4$, (c) H$_2^+$, and (d) H$_2$H$_2$.
Dashed black lines are the exact curves. Dashed brown lines denote the results computed from uniform gas LDA and blue solid lines denote the results computed from KSR-global.
}
\end{figure}

Figure~\ref{fig:pes} shows the dissociation curve of H$_2$, H$_4$, H$_2^+$ and H$_2$H$_2$. Curves of KSR-global (blue) are computed from the neural XC functional trained from the full KSR with $N_\mathrm{train}=8$ (four H$_2$ molecules and four H$_4$ molecules) in Figure~\ref{fig:generalization_to_20}. 
KSR fits H$_2$ and H$_4$ well even in the stretched limit. KSR perfectly predicts H$_2^+$ because of self-interaction gate in the neural XC functional. Although H$_2$H$_2$ has never been exposed to
the model during training, KSR performs well at small distances ($R<3$) and at large distances ($R>8$) but slightly overbinds around $R=5$.

\section{Neural networks}

\subsection{Architecture}

\begin{figure}[htb]
\includegraphics[width=0.7\columnwidth]{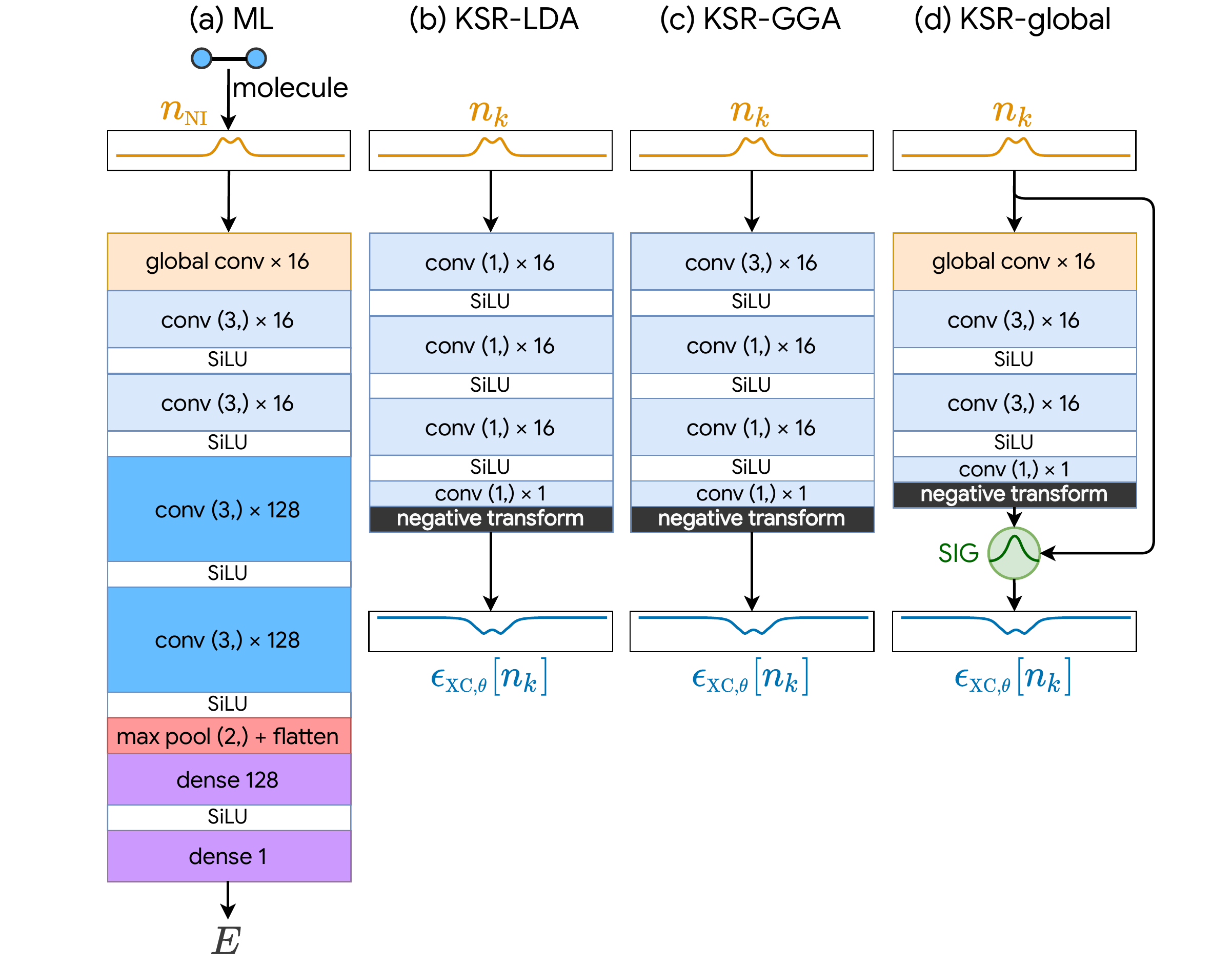}
\caption{\label{fig:models} Model architectures of (a) the ML model that directly predicts energy $E$ from geometry, (b) the neural LDA with KSR, (c) the neural GGA with KSR, and (d) the neural global functional.
}
\end{figure}

Figure~\ref{fig:models} illustrates the model architectures used in Figure~1. In the ML model that directly predicts energy from geometry (Figure~\ref{fig:models}(a)), we first solve Eq.~\ref{eq:non_interacting_eqns} to obtain a density for a particular molecular geometry. We use this density as a smooth representation of the geometry. The first few layers (global conv-conv-SiLU-conv-SiLU) are identical to the KSR-global in Figure~\ref{fig:models}(d). Next, we use convolution layers with 128 channels and dense layers to increase the capacity of the model. Finally, a dense layer with a single unit outputs the scalar $E$.

The KSR-LDA and KSR-GGA approaches do not have the global convolution layer because the use of global information violates the local and semi-local approximation. The first layer of KSR-LDA is a convolution with filter size 1. It mimics the physics of the standard LDA approach by mapping the density value to the XC energy density at the same point, $\epsilon\lda\xctheta:\mathbb{R}^1\to\mathbb{R}^1$. KSR-GGA uses a convolution layer with filter size 3 to map the density values of three nearby points to the XC energy density at the center point, $\epsilon\gga\xctheta:\mathbb{R}^3\to\mathbb{R}^1$. The XC energy density in the entire space is also computed pointwise, $\epsilon\xc=\big\{\epsilon\gga\xctheta[n(x_{-1},x_0,x_1)], \ldots, \epsilon\gga\xctheta[n(x_{m-2},x_{m-1},x_m)]\big\}\in \mathbb{R}^m$. The remaining network structure of KSR-LDA and KSR-GGA is identical to KSR-global, except for self-interaction gate (SIG).

\subsection{Layers}

\subsubsection{Convolution}
Filter weights in 1D convolution are initialized by He normal initialization~\cite{he2015delving}. The stride is one. The edges are padded with zero to ensure that the size of the output spatial dimension is the same as the size of the unpadded input spatial dimension. There is no bias term.

\subsubsection{Global convolution}
Global convolution contains multiple channels to capture the interaction in different scales. The operation in each channel is 
\begin{equation}
    G(n(x),\xi_p) = \frac{1}{2\xi_p}\int dx' n(x') \exp(-|x-x'|/\xi_p).
\end{equation}
where $\xi_p$ is trainable and controls the scale of the interaction. We parameterize $\xi_p=a + (b-a)\cdot\sigma(\eta_p)$ using the sigmoid function $\sigma(x)=1/(1+\exp(-x))$ to bound $\xi_p\in(a,b)$. $\eta_p$ is initialized using the normal distribution $\mathcal{N}(0,10^{-4})$. For the 16-channel global convolution layer used in this work, we have $\eta_1\equiv0$ to preserve the input density and the rest $\eta_p\in(0.1, \kappa^{-1})$ are trainable.

\subsubsection{Dense}
The dense layer is only used in the ML model (Figure~\ref{fig:models}(a)). The weights are initialized using Glorot normal initialization~\cite{glorot2010understanding} and bias terms are initialized using the normal distribution $\mathcal{N}(0,10^{-4})$.

\subsection{Checkpoint selection}
Each calculation is repeated with 25 random seeds. The model is trained by L-BFGS~\cite{liu1989limited} implemented in SciPy~\cite{scipy} \verb|scipy.optimize.fmin_l_bfgs_b| with \verb|factr=1,m=20,pgtol=1e-14|.
Parameter checkpoints are saved every 10 steps until L-BFGS stops. The optimal checkpoint is the checkpoint with the lowest average energy error per electron on validation sets,
$\mathbb{E}_{\{\mathcal{S}_\mathrm{val}\}} \mathbb{E}_{\mathcal{M}\in \mathcal{S}_\mathrm{val}} |(E - E\dmrg)|/ N_e$,
where $E$ is the final energy from KS calculations, $E\dmrg$ is the exact energy, and $N_e$ is the number of electrons. The validation sets $\{\mathcal{S}_\mathrm{val}\}$ and the molecules $\mathcal{M}$ in each sets are listed in Table~\ref{table:validation_molecules}.

\section{Training a neural XC functional without KS regularization}
\citet{schmidt2019machine} proposed a neural XC functional that can be used in a KS self-consistent calculation in the inference stage. Unlike our work, which trains the network through KS calculations, they train the network in a single-step. The training set contains 12800 molecules and the validation set contains 6400 molecules. Here, the molecules are exact solutions of one-dimensional two-electron problems in the external potential of up to three random nuclei (Equation~4 in \citet{schmidt2019machine}). The exact $v\xc$ for each molecule is computed by an inverse KS method. They input exact ground state density and train the network to predict XC energy per length $e\xc$ that minimizes the loss function: a weighted combination of the mean square errors (MSE) of the XC energy, XC potential, its numerical spatial derivative, and the difference between the XC energy and the integral over the potential (Equation~5 in \citet{schmidt2019machine}),
\def\MSE{\, \text{MSE}}
\begin{equation}
    L(\theta)=\alpha \MSE(E\xc)+\beta \MSE(v\xc)+\gamma \MSE\Big(\frac{d v\xc(x)}{dx}\Big) + \delta \MSE\Big(E\xc-\int dx \, v\xc(x) n(x)\Big),
\end{equation}
where $\alpha=1.0$, $\beta=100.0$, $\gamma=10.0$, and $\delta=1.0$ is the weights used in \citet{schmidt2019machine}.

\begin{figure}[htb]
\includegraphics[width=0.6\columnwidth]{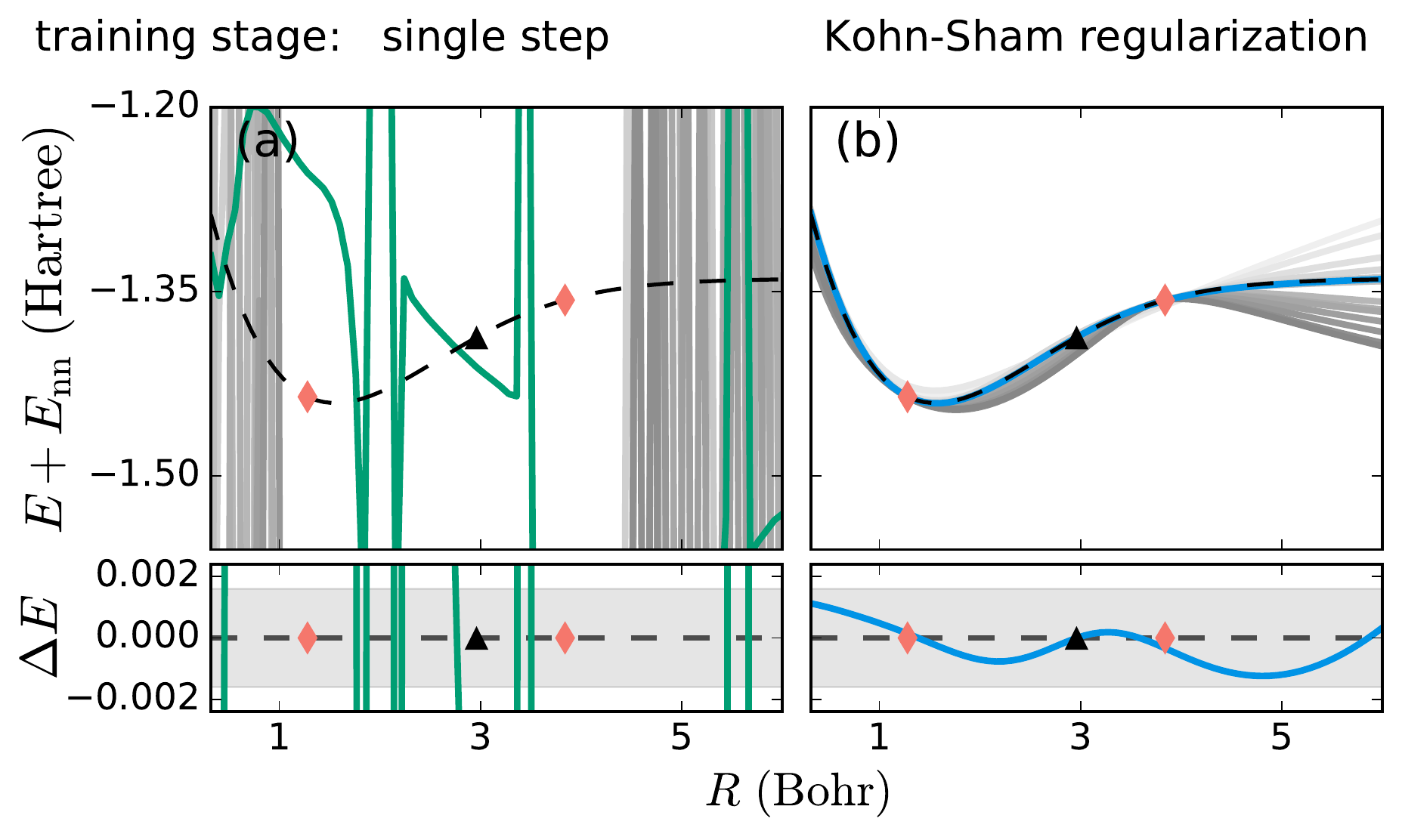}
\caption{\label{fig:with_without_ksr}
Training the neural XC functional using (a) single-step and (b) Kohn-Sham regularization. Both functionals use Kohn-Sham self-consistent calculations in the inference stage.
}
\end{figure}

It is natural to pose the question: does the generalization from the two H$_2$ training molecules in Figure~1 result from using KS self-consistent calculations in the inference stage rather than the training stage? This is a reasonable concern because the XC energy is usually a small portion of the total energy. To justify this concern, we first use inverse KS to get the exact $v\xc$ on the two H$_2$ molecules used in the H$_2$ experiment. Then, we take the model architecture~\footnote{The only difference is that this model predicts $\epsilon\xc[n](x)$ rather than $e\xc[n](x)$ in \citet{schmidt2019machine}. The relation between them is $e\xc[n](x)=\epsilon\xc[n](x)\cdot n(x)$} and loss function in \citet{schmidt2019machine} and attempt to learn the entire dissociation curve of H$_2$ from two molecules. Figure~\ref{fig:with_without_ksr} compares the results from KS self-consistent calculations using functionals trained on (a) single-step and (b) Kohn-Sham regularization. It is not surprising that even though both approaches use KS self-consistent calculations in the inference stage, the model trained on a single-step fails to generalize in the small training set limit ($1/6400$ training set size to the original paper). The neural XC functional is a many-to-many mapping, which is very hard to learn with limited data. Moreover, KS self-consistent calculations start with an initial density that is not the exact ground state density. It is clearly out of the interpolation region for the model that has only seen exact densities of two molecules.

We would like to emphasize that this comparison aims to show that using KS calculations in {\it training} -- Kohn-Sham regularizer -- is crucial to the generalization.
A single-step model could work well as reported in \citet{schmidt2019machine} with a larger training set and exact $v\xc$. 

\section{Training a neural XC functional with ``weaker'' Kohn-Sham regularization}

\begin{figure}[htb]
\includegraphics[width=0.8\columnwidth]{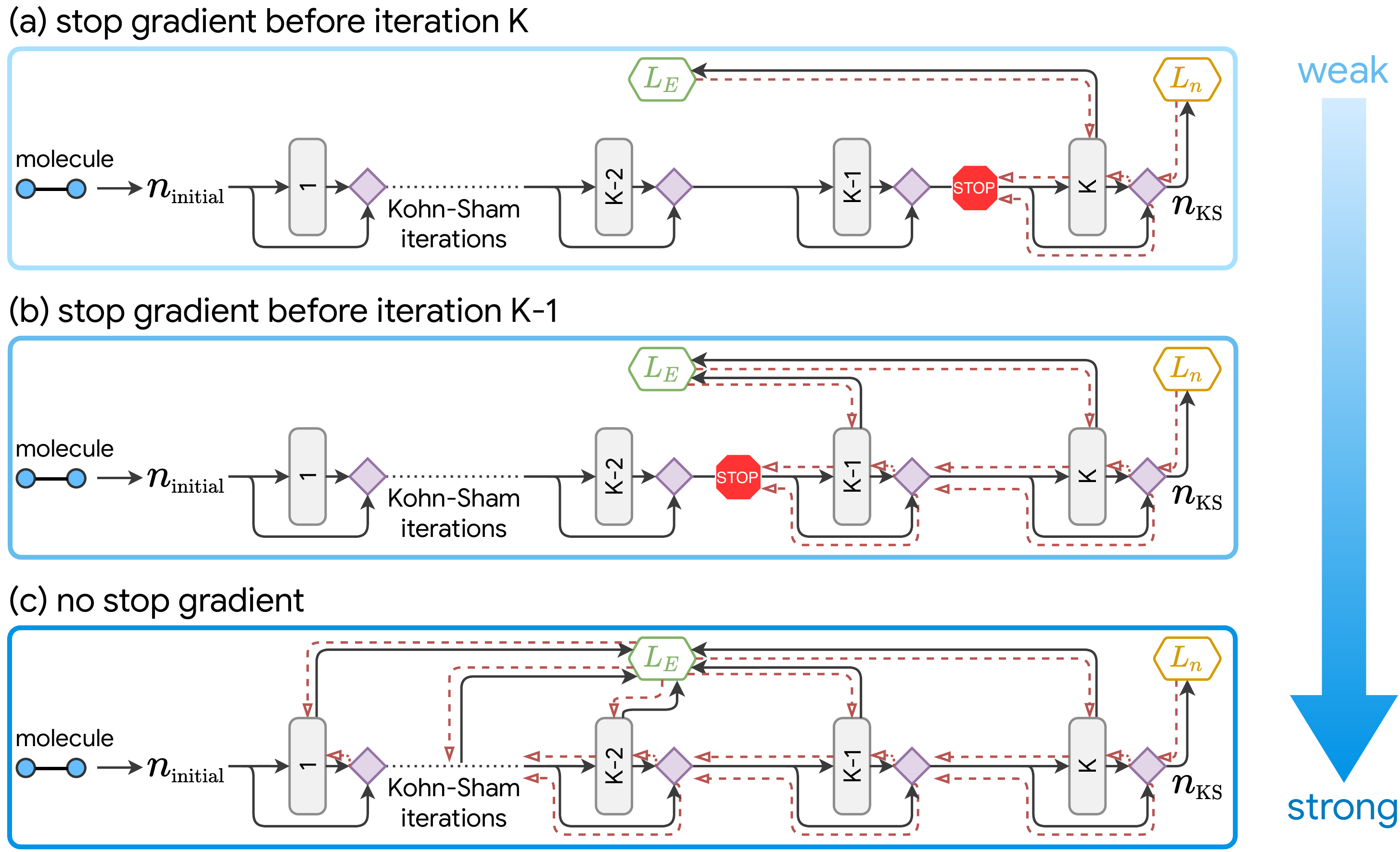}
\caption{\label{fig:stop_gradient}
Computational graph with different KSR strength. (a) stop gradient before iteration $K$. (b) stop gradient before iteration $K-1$. (c) no stop gradient. This is the same computation graph used in the main text.
}
\end{figure}

\begin{figure}[htb]
\includegraphics[width=0.6\columnwidth]{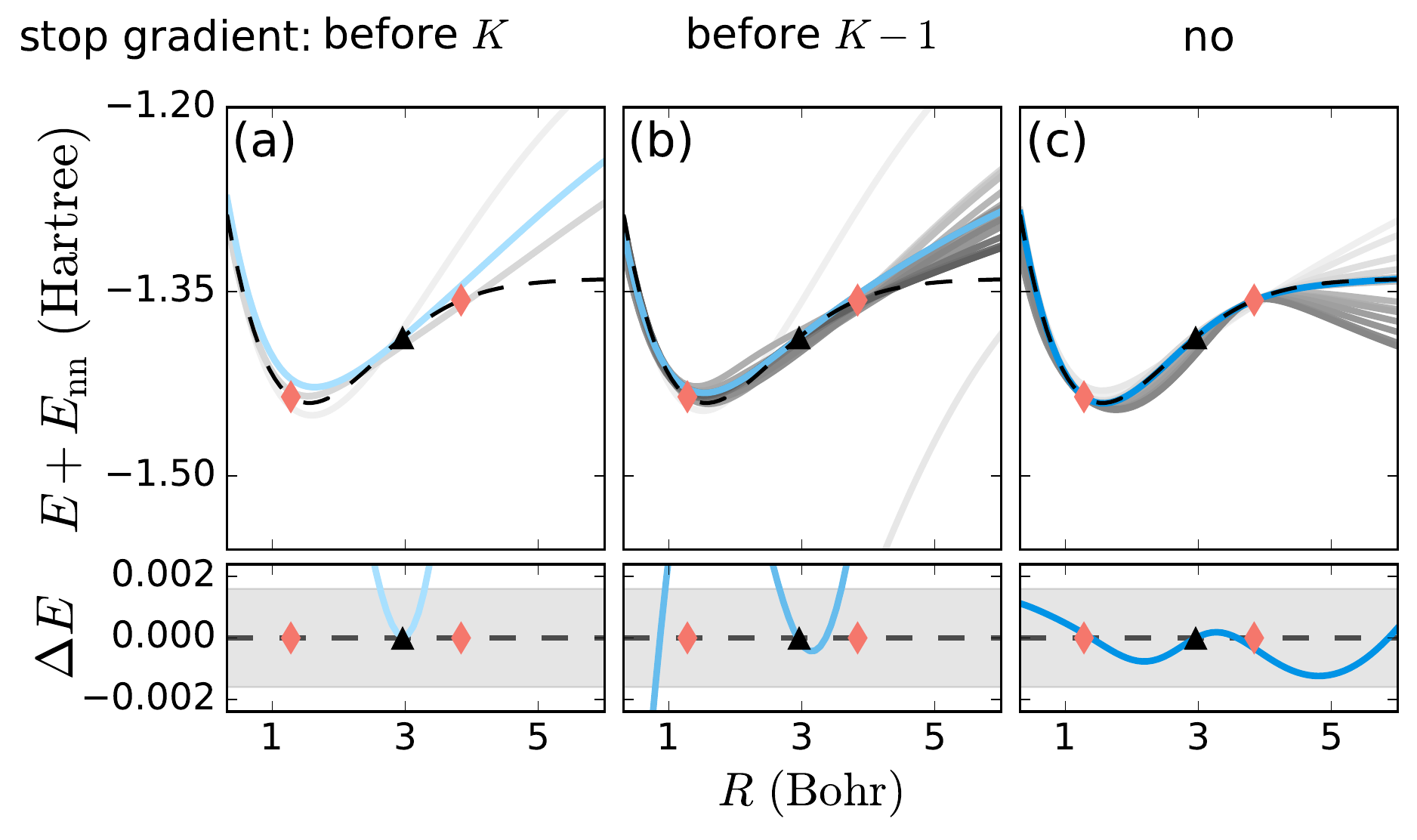}
\caption{\label{fig:with_weaker_ksr}
H$_2$ dissociation curves trained from two molecules (red diamonds) with a corresponding computational graph shown in Figure~\ref{fig:stop_gradient}.
}
\end{figure}

Unlike other methods that build physics prior knowledge to the model through constraint, KSR ``augments'' densities for the model during training. Thus, there is no single coefficient to explicitly control the strength of the regularization. A straightforward idea to control the KSR strength is to change the total number of iterations $K$ in the KS self-consistent calculations. However, a small $K$ may not be sufficient to converge KS calculations. Thus it is ambiguous to understand whether the worse performance is from weaker regularization or unconverged KS calculations. Here we design an approach to control the strength of KSR by stopping the gradient flow in the backpropagation and keeping $K$ fixed.

{\it Stop gradient} is a common operation in differentiable programming. It acts as an identity in the forward pass so it does not affect the KS calculations. In the backpropagation, it sets the gradient passing through it to zero. 
As shown in Figure~\ref{fig:stop_gradient}, we stop gradient before a certain KS iteration $k=k^*$ so all the previous iterations $k<k^*$ have no access to the gradient information. Since the gradient may still flow into the iteration $k$ from $L_E$ through its energy output $E_k$, we also stop the gradient on $E_k$ for $k<k^*$. To simplify the graph, we remove the arrows between $E_k$ to $L_E$ for $k<k^*$.
In Figure~\ref{fig:stop_gradient}(a), the neural XC functional is updated only from the gradient information flowing through the final iteration.
(b) is similar to (a) but has access to the gradient flowing through the last two iterations.
No stop gradient is applied to (c) and it is identical to the computational graph we used in the main text.
We repeat the same experiment in Figure~1. Figure~\ref{fig:with_weaker_ksr} shows the H$_2$ dissociation curves trained with three stop gradient setting in Figure~\ref{fig:stop_gradient}.
In Figure~\ref{fig:with_weaker_ksr}(a), L-BFGS converges quickly as there is no sufficient gradient information for training. By including the gradient information in the $K-1$-th iteration, the distribution of the dissociation curves predicted by the model during training get closer to the true curve in (b). For comparison, we place the distribution of dissociation curves from model without stop gradient in (c), previously shown in Figure~1(d), where the physics of the true dissociation curve is captured. 

\bibliography{references}